# Consistent formalism for the momentum of electromagnetic waves in lossless dispersive metamaterials and the conservation of momentum

**Yingran He[1], Jianqi Shen[1] and Sailing He[1,2*]**

[1] *Centre for Optical and Electromagnetic Research, JORCEP[KTH-ZJU Joint Research Center of Photonics], East Building #5, Zijingang campus, Zhejiang University (ZJU), Hangzhou 310058, China*

[2] *Department of Electromagnetic Engineering, School of Electrical Engineering, Royal Institute of Technology (KTH), S-100 44 Stockholm, Sweden*

** Corresponding author: sailing@kth.se*

**Abstract-** A new formalism for electromagnetic and mechanical momenta in a metamaterial is developed by means of the technique of wave-packet integrals. The medium has huge mass density and can therefore be regarded as almost stationary upon incident electromagnetic waves. A clear identification of momentum density and momentum flow, including their electromagnetic and mechanical parts, is obtained by employing this formalism in a lossless dispersive metamaterial (including the cases of impedance matching and mismatching with vacuum). It is found that the ratio of the electromagnetic momentum density to the mechanical momentum density depends on the impedance and group velocity of the electromagnetic wave inside the metamaterial. One of the definite results is that both the electromagnetic momentum and the mechanical momentum in the metamaterial are in the same direction as the energy flow, instead of in the direction of the wave vector. The conservation of total momentum is verified. In addition, the law of energy conservation in the process of normal incidence is also verified by using the wave-packet integral of both the electromagnetic energy density and the electromagnetic power density, of which the latter is caused by the interaction between the induced (polarized) currents and the electromagnetic wave.

## 1. INTRODUCTION

The expressions/definitions for the electromagnetic momentum density and momentum stress tensor in vacuum have been clearly identified and widely accepted [1]. However, the expressions for the electromagnetic momentum density and momentum stress tensor in a medium have been debated for over a century. The earliest expressions for momentum density in a dielectric medium were proposed by Abraham [2] as $\vec{g} = \varepsilon_0 \mu_0 \vec{E} \times \vec{H}$ and by Minkwoski [3] differently as $\vec{g} = \vec{D} \times \vec{B}$. Proceeding theoretical and experimental works have been carried out [4-14], trying to solve this controversy. Early experiment by Jones and Richards found that the radiation pressure on a mirror immersed in a refracting liquid medium is proportional to the refractive index of the liquid [4]. Theoretical analyses by Gordon [5] and Peierls [6] have concluded that: (a) Abraham's momentum gives the electromagnetic part momentum of waves in the medium, but excludes the mechanical part momentum carried by the medium itself; (b) Minkwoski's momentum is actually a pseudo-momentum, which results from the invariance of physical laws with respect to the displacement of medium coordinates; This can be illustrated more clearly using both spatial (Eulerian) and material (Lagrangian) coordinates [7, 8]. (c) The Minkwoski's momentum is mathematically useful in determining the radiation pressure exerted on a mirror by incident light, although the radiation pressure origins from both electromagnetic momentum and mechanical momentum of waves. Recent studies focus on the calculation of momentum of waves in different media, from ordinary dielectrics [9] to magnetic media [10, 13], and even to dispersive media [11, 12, 14]. A recent review paper can also be found for an overview on this subject [15].

Metamaterials with simultaneously negative permittivity and negative permeability may achieve unprecedented electromagnetic properties, such as negative refraction [16, 17] and potentially perfect imaging [18]. However, far less attention has been paid to the investigation of the momentum for electromagnetic waves in metamaterials [12, 19, 20]. Scalora et al. [12] showed that the Lorentz force in a metamaterial was consistent with the rate of change of Abraham's momentum, but they did not obtain the ratio of the mechanical part momentum to the electromagnetic part momentum for electromagnetic waves. Kemp et al. [19] claimed that the momentum flow in a left-handed material was opposite to the power flow direction, but we believe this conclusion still requires further argument/research. Veselago has noticed the incompatibility between photon momentum flow and energy transfer for electromagnetic waves in metamaterials, and assumed that the momentum inside a metamaterial is in the same direction as the wave vector [20], which is not correct as we show in the present paper.

## 2. MODELING

It has been pointed out that a complete expression for the momentum in a medium needs to take both the electromagnetic part momentum and the mechanical part momentum into account [15]. The continuity equations for them are

$$\frac{\partial}{\partial t}\vec{g}_{\text{em}} + \nabla \bullet \overline{\overline{T}}_{\text{em}} = -\vec{f}_{\text{mech}} \tag{1}$$

$$\frac{\partial}{\partial t}\vec{g}_{\text{mech}} + \nabla \bullet \overline{\overline{T}}_{\text{mech}} = \vec{f}_{\text{mech}} \tag{2}$$

where $\vec{g}_{\text{em}}$, $\overline{\overline{T}}_{\text{em}}$, $\vec{g}_{\text{mech}}$ and $\overline{\overline{T}}_{\text{mech}}$ represent the momentum density and momentum flow density tensor (momentum stress tensor) for the electromagnetic part and the mechanical part, respectively.

The sum of the above two equations yields the expression for the conservation of the total momentum

$$\frac{\text{d}}{\text{d}t}\int_V \vec{g}_{\text{tot}} dV + \oint_S d\vec{S} \bullet \overline{\overline{T}}_{\text{tot}} = 0 \tag{3}$$

Here, $\vec{g}_{\text{tot}}$ and $\overline{\overline{T}}_{\text{tot}}$ represent the total momentum density and momentum flow density tensor inside the medium, respectively.

By directly applying Maxwell's equations, we can obtain the following expressions for the momentum density of electromagnetic field $\vec{g}_{\text{em}}$, the generalized Lorentz force density $\vec{f}_{\text{mech}}$, and the electromagnetic momentum flow density $\overline{\overline{T}}_{\text{em}}$,

$$\vec{g}_{\text{em}} = \varepsilon_0 \mu_0 \vec{E} \times \vec{H} \tag{4}$$

$$\vec{f}_{\text{mech}} = -(\nabla \bullet \vec{P})\vec{E} - \mu_0(\nabla \bullet \vec{M})\vec{H} + \mu_0(\partial \vec{P} / \partial t) \times \vec{H} - \mu_0 \varepsilon_0 (\partial \vec{M} / \partial t) \times \vec{E} \tag{5}$$

$$\overline{\overline{T}}_{\text{em}} = \frac{1}{2}(\varepsilon_0 \vec{E} \cdot \vec{E} + \mu_0 \vec{H} \cdot \vec{H})\overline{\overline{I}} - (\varepsilon_0 \vec{E}\vec{E} + \mu_0 \vec{H}\vec{H}) \tag{6}$$

where $\vec{D} = \varepsilon_0 \vec{E} + \vec{P}$ and $\vec{B} = \mu_0(\vec{H} + \vec{M})$ have been substituted. The momentum density of electromagnetic field is in Abraham's form, and the four terms in Eq. (5) represent the force on the electrically polarized charge by the electric field, the force on the magnetically polarized charge by the magnetic field, the force on the electrically polarized current by the magnetic field and the force on the magnetically polarized current by the electric field, respectively. The electromagnetic momentum flow density tensor relates to $\vec{E}$ and $\vec{H}$, rather than $\vec{D}$ and $\vec{B}$.

It can be verified that $\vec{f}_{\text{mech}} \delta V$ equals the mechanical momentum increase rate inside the macroscopic small volume $\delta V$ [14]. In other words, the generalized Lorentz force plays the role of converting the electromagnetic momentum into the mechanical momentum, while the total momentum is conserved. Due to the action of the generalized Lorentz force, the unit cell of the medium will gain certain mechanical momentum and, therefore, be set into motion.

In general, motion of unit cells inside the medium will not only give rise to mechanical momentum and mechanical momentum flow, but also have an impact on the constitutive relation of the medium [8]. Therefore, the mechanical properties of a dielectric material have to be specified, in order to describe how it responses to the action of the generalized Lorentz force. To simplify the calculation, in this paper we aim at deriving an explicit expression for momentum in a homogenous isotropic medium, which is constituted by unit cells with very large mass, i.e., huge mass density $M$, such that the motion of dielectrics is sufficiently small and the medium remains almost stationary and consequently the constitutive relation of the medium remains the same upon electromagnetic wave incidence.

Although velocity $\vec{v}$ of the mass center of the unit cell is infinitesimal as a result of huge mass density $M$, their product $M\vec{v}$, representing the accompanied mechanical momentum density $\vec{g}_{\text{mech}}$, is a comparable quantity with respect to electromagnetic momentum density $\vec{g}_{\text{em}}$. Note that the mechanical momentum flow $\overline{\overline{T}}_{\text{mech}}$, of the order $(M\vec{v})\vec{v}$, is still negligibly small compared with electromagnetic momentum flow $\overline{\overline{T}}_{\text{em}}$.

Based on the above analysis, we can obtain a clear physical interpretation of the momentum transfer and conversion for propagating waves inside our medium model. When the leading edge of a travelling wave passes though a macroscopic small volume, the volume will not only gain a momentum carried by the electromagnetic wave itself, but also acquire a mechanical momentum carried by the infinitesimal movement of the huge mass, namely $M\vec{v}$, due to the presence of the generalized Lorentz force. The force will decrease gradually as the electromagnetic field grows inside this volume and eventually vanish when the electromagnetic field becomes time-harmonic. After that, both the electromagnetic momentum density and the mechanical momentum density of this small volume reach some steady values, and a constant momentum flow is carried by the electromagnetic field, which transmits the required momentum to establish a time-harmonic field at the new leading edge of the propagating wave.

The additional mechanical momentum density $M\vec{v}$ can be evaluated by integrating the Lorentz force over the time for establishing a time-harmonic field. For a plane wave in a dispersion-free non-magnetic dielectric with refractive index $n$, it has been shown that [6, 9]

$$g_{\text{mech}} = \frac{n^2 - 1}{2} g_{\text{em}} \tag{7}$$

In this paper, we would like to focus on the momentum density and momentum flow in a metamaterial, which is intrinsically dispersive. Inside a dispersive medium, the medium's response to the driving of an incident field is time-dependent in a causal way, i.e., its response at time $\tau_0$ is related not only to the driving field at time $\tau_0$, but also to the driving field at times previous to $\tau_0$. To treat this dispersive property, it is convenient to construct a wave packet, and investigate the increase of the mechanical momentum along with the establishment of this wave packet inside the medium. In the limit that the wave packet is long enough, a correct expression for momentum density can be derived (see [11] for a dispersive dielectric case). It is interesting to notice that the momentum flow density tensor is independent of dispersion.

## 3. MOMENTUM DENSITY, MOMENTUM FLOW AND MOMENTUM CONSERVATION

Since the momentum of electromagnetic waves in vacuum is unambiguous, the fundamental physical law of momentum conservation can be used to check the validity of our modeling theory for the case of normal incidence of plane waves from vacuum into a semi-infinite medium. In this paper the metamaterial is assumed to be lossless for the sake of simplicity and clarity.

### 3.1 Momentum for the case of matched impedance

Now we study the case when a plane wave normally impinges from vacuum on a semi-infinite metamaterial with negative relative permittivity $\varepsilon_{\text{r}}$ and negative relative permeability $\mu_{\text{r}}$.

A metamaterial is always dispersive. First we assume $\varepsilon_{\text{r}}$ and $\mu_{\text{r}}$ have the same dispersion, i.e., $\varepsilon_{\text{r}}(\omega)=\mu_{\text{r}}(\omega)$, so that its impedance matches with the vacuum impedance for all frequencies to ensure total transmission. In order to deal with a dispersive material, it is convenient to construct a wave packet by using two spectrum components whose frequencies are close to each other, similar to the treatment of energy density inside a dispersive dielectric medium [1].

In vacuum, the incident field components are

$$\begin{cases} E_x^{\text{i}} = E_0 \sin(k_0^a z - \omega_a t) - E_0 \sin(k_0^b z - \omega_b t) \\ H_y^{\text{i}} = \sqrt{\varepsilon_0 / \mu_0} E_0 \sin(k_0^a z - \omega_a t) - \sqrt{\varepsilon_0 / \mu_0} E_0 \sin(k_0^b z - \omega_b t) \end{cases} \tag{8}$$

where $k_0^\nu$ is the vacuum wave vector corresponding to $\omega_\nu$ with $\nu = a, b$.

In the metamaterial, the transmitted field components are

$$\begin{cases} E_x^{\text{t}} = E_0 \sin(k_a z - \omega_a t) - E_0 \sin(k_b z - \omega_b t) \\ H_y^{\text{t}} = \sqrt{\varepsilon_0 / \mu_0} E_0 \sin(k_a z - \omega_a t) - \sqrt{\varepsilon_0 / \mu_0} E_0 \sin(k_b z - \omega_b t) \end{cases} \tag{9}$$

where $k_\nu = -\sqrt{\varepsilon_{\text{r}}^\nu \mu_{\text{r}}^\nu} k_0^\nu = \varepsilon_{\text{r}}^\nu k_0^\nu$, $\varepsilon_{\text{r}}^\nu$ and $\mu_{\text{r}}^\nu = \varepsilon_{\text{r}}^\nu$ are the wave vector, relative permittivity and relative permeability at frequency $\omega_\nu$ with $\nu = a, b$. In the limit of $\omega_a \to \omega_b \to \omega$, the wave vector, relative permittivity and relative permeability corresponding to the central frequency $\omega$ will be denoted by $k$, $\varepsilon_\mathrm{r}$ and $\mu_\mathrm{r}$, respectively.

Now we focus on the momentum conversion process of this wave packet during a time interval [0, $\tau$]. $\tau = 2\pi/(\omega_a - \omega_b)$ corresponds to the period of this wave packet envelope.

At $t$=0, the incident wave packet which locates at $z\in$[-$c\tau$, 0] has the following momentum

$$\begin{aligned}
P_\mathrm{vac} &= \int_{-c\tau}^{0} (g_\mathrm{em})_\mathrm{in}\, dz = \int_{-c\tau}^{0} \varepsilon_0 \mu_0 E_x^\mathrm{i} H_y^\mathrm{i} dz \\
&= \frac{\varepsilon_0 E_0^2}{c}\int_{-c\tau}^{0} \left(\sin(k_0^a z - \omega_a t) - \sin(k_0^b z - \omega_b t)\right)\left(\sin(k_0^a z - \omega_a t) - \sin(k_0^b z - \omega_b t)\right) dz \\
&= \frac{\varepsilon_0 E_0^2}{c}\int_{-c\tau}^{0} \left\{ \begin{aligned} &1 - \frac{\cos(2k_0^a z - 2\omega_a t)}{2} - \frac{\cos(2k_0^b z - 2\omega_b t)}{2} \\ &+ \cos\left[(k_0^a + k_0^b)z - (\omega_a + \omega_b)t\right] - \cos\left[(k_0^a - k_0^b)z - (\omega_a - \omega_b)t\right] \end{aligned} \right\} dz
\end{aligned} \tag{10}$$

where $c$ is the optical velocity in vacuum.

It is noted that the one-packet spatial integral of the highly-oscillating terms $\cos(2k_0^a z - 2\omega_a t)$, $\cos(2k_0^b z - 2\omega_b t)$ and $\cos\left[(k_0^a + k_0^b)z - (\omega_a + \omega_b)t\right]$ would vanish when we take the limit of $\omega_a \to \omega_b \to \omega$, and that the wave-packet spatial integral of the slowly-oscillating term $\cos\left[(k_0^a - k_0^b)z - (\omega_a - \omega_b)t\right]$ is also zero, i.e.,

$$\begin{aligned}
&-\frac{\varepsilon_0 E_0^2}{c}\int_{-c\tau}^{0} \cos\left((k_0^a - k_0^b)z - (\omega_a - \omega_b)t\right) dz \\
&= -\frac{\varepsilon_0 E_0^2}{c}\frac{1}{k_0^a - k_0^b}\left[\sin\left((k_0^a - k_0^b)(c\tau) + (\omega_a - \omega_b)t\right) - \sin\left((\omega_a - \omega_b)t\right)\right] \\
&= 0
\end{aligned}$$

Therefore, the incoming momentum is easily found as

$$P_\mathrm{vac} = \varepsilon_0 E_0^2 \tau \tag{11}$$

At $t$=$\tau$, this wave packet will completely penetrate into the metamaterial at the location $z\in$[0, $v_\mathrm{g}\tau$], where $v_\mathrm{g} \equiv \frac{\mathrm{d}\omega}{\mathrm{d}k} = \frac{\omega_a - \omega_b}{k_a - k_b}$ is the group velocity. In this case, $v_\mathrm{g} \equiv \frac{\mathrm{d}\omega}{\mathrm{d}k} = \frac{\mathrm{d}\omega}{\mathrm{d}(\omega\varepsilon_\mathrm{r}/c)} = \frac{c}{\mathrm{d}(\omega\varepsilon_\mathrm{r})/\mathrm{d}\omega}$ is always positive since a normal dispersion, i.e., $\mathrm{d}\varepsilon_\mathrm{r}/\mathrm{d}\omega > 0$, is assumed in our investigation. The total momentum inside the metamaterial consists of two parts, the electromagnetic part and the mechanical part. Similar to the calculation of $P_\mathrm{vac}$, electromagnetic part momentum $P_\mathrm{em}$, and spatially averaged momentum density $g_\mathrm{em}$

are

$$P_{\text{em}} = \int_0^{v_g\tau} g_{\text{em}} dz = \int_0^{v_g\tau} \varepsilon_0\mu_0 E_x^{\text{t}} H_y^{\text{t}} dz = \varepsilon_0 E_0^2 v_g \tau / c \tag{12}$$

$$g_{\text{em}} = P_{\text{em}} / v_g\tau = \varepsilon_0 E_0^2 / c \tag{13}$$

The mechanical part of the momentum is produced by the action of Lorentz force, so that it can be calculated by integrating the Lorentz force exerted on this wave packet, i.e.,

$$P_{\text{mech}} = \int_0^{\tau}\int_0^{v_g t} f_{\text{mech}}(z,t) dz dt \tag{14}$$

In this situation, the $z$ component of Lorentz force, $f_z^{\text{mech}}$, turns out to be

$$\begin{aligned} f_z^{\text{mech}} &= \mu_0\varepsilon_0[-\omega_a(\varepsilon_r^a - 1)E_0\cos(k_a z - \omega_a t) + \omega_b(\varepsilon_r^b - 1)E_0\cos(k_b z - \omega_b t)]H_y^{\text{t}} \\ &+ \mu_0\varepsilon_0[-\omega_a(\mu_r^a - 1)\sqrt{\varepsilon_0/\mu_0}E_0\cos(k_a z - \omega_a t) + \omega_b(\mu_r^b - 1)\sqrt{\varepsilon_0/\mu_0}E_0\cos(k_b z - \omega_b t)]E_x^{\text{t}} \end{aligned} \tag{15}$$

After calculation (see Appendix), it is found that the mechanical momentum is

$$P_{\text{mech}} = \varepsilon_0\mu_0\sqrt{\varepsilon_0/\mu_0}E_0^2\tau[(\varepsilon_r^a - 1)\omega_a - (\varepsilon_r^b - 1)\omega_b]/(k_a - k_b) \tag{16}$$

In the limit of $\omega_a \to \omega_b \to \omega$, we obtain the following mechanical momentum inside this wave packet

$$P_{\text{mech}} = \varepsilon_0 E_0^2\tau\left(\frac{\text{d}(\omega\varepsilon_r)}{\text{d}\omega} - 1\right) / \frac{\text{d}(\omega\varepsilon_r)}{\text{d}\omega} = \varepsilon_0 E_0^2 \tau(1 - v_g / c) \tag{17}$$

and subsequently the spatially averaged mechanical momentum density is obtained

$$g_{\text{mech}} = P_{\text{mech}} / v_g\tau = \varepsilon_0 E_0^2 (1/v_g - 1/c) \tag{18}$$

It follows from Eqs. (11), (12) and (17) that $P_{\text{vac}} = P_{\text{em}} + P_{\text{mech}}$, which means that the law of momentum conservation holds during the period when the electromagnetic wave penetrates into the metamaterial from vacuum (see Fig. 1)

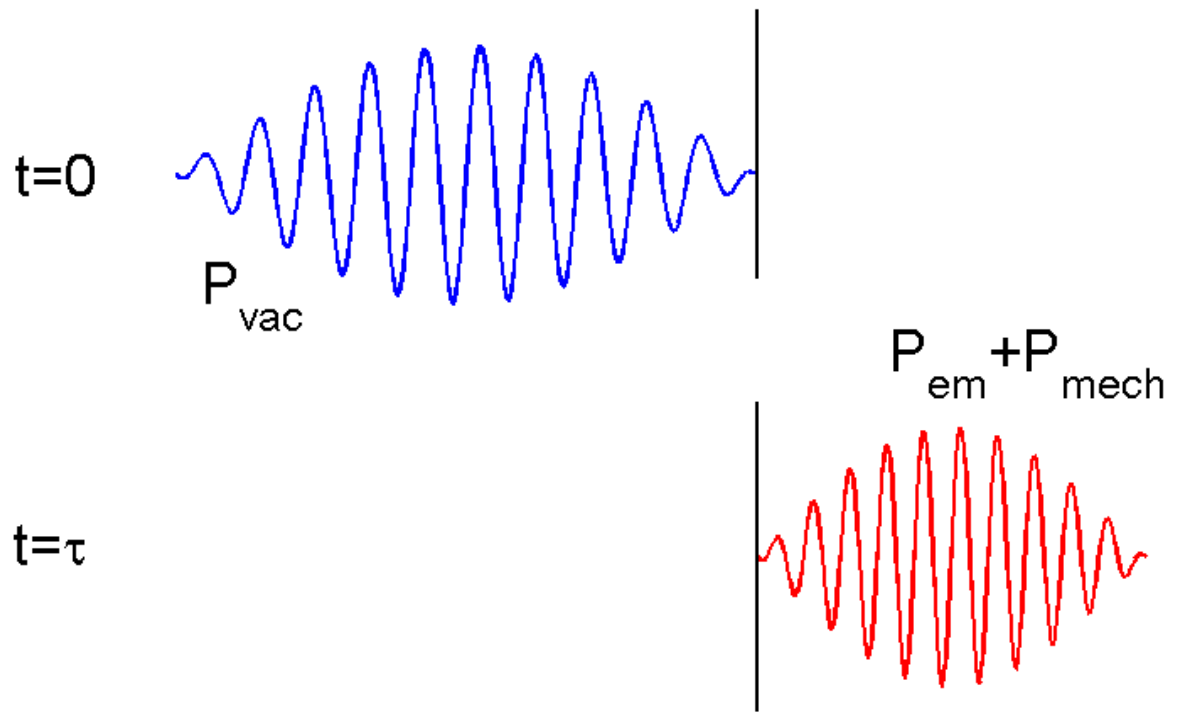

Figure 1. Illustration of momentum conservation at the interface between vacuum and an impedance-matched metamaterial. The incident wave packet (in blue) carries momentum $P_{\text{vac}}$, while the transmitted wave packet (in red) carries momentum $P_{\text{em}}+P_{\text{mech}}$.

Since a total momentum $P_{\text{em}} + P_{\text{mech}}$ is transmitted into the metamaterial from the vacuum during a time interval [0, $\tau$], the total momentum flow in metamaterial is

$$p_z^{\mathrm{tot}} = (P_{\mathrm{em}} + P_{\mathrm{mech}}) / \tau = \varepsilon_0 E_0^2 \tag{19}$$

This result is in agreement with the theoretical prediction by the momentum flow density tensor $T_{zz}^{\mathrm{tot}} = T_{zz}^{\mathrm{em}} = \varepsilon_0 E_0^2$ in the metamaterial, according to Eq. (6).

In a brief summary, for a $+z$ propagating plane wave in a metamaterial, our results show that

$$\begin{cases} k = -\sqrt{\varepsilon_{\mathrm{r}}\mu_{\mathrm{r}}}\, k_0 = \varepsilon_{\mathrm{r}} k_0 \\ S_z = \sqrt{\varepsilon_0 / \mu_0}\, E_0^2 \\ p_z^{\mathrm{tot}} = \varepsilon_0 E_0^2 \\ g_{\mathrm{em}} = S_z / c^2 \\ g_{\mathrm{mech}} = (c / v_{\mathrm{g}} - 1) g_{\mathrm{em}} \end{cases} \tag{20}$$

It is worth noticing that, inside the metamaterial the wave vector is in $-z$ direction, but the momentum density and momentum flow (both the electromagnetic part and mechanical part) of the plane wave is in $+z$ direction (as the energy flow). Therefore, we conclude that, within the classical electrodynamics framework, the total momentum of electromagnetic waves in a metamaterial has the same direction as the energy flow, instead of as the wave vector [19, 20].

**3.2 Momentum for the case of mismatched impedance**

In this situation, we would like to perform a calculation on a more general case, i.e., $\varepsilon_{\mathrm{r}}(\omega) \neq \mu_{\mathrm{r}}(\omega)$ to show that our formalism is applicable for more general situations.

The components for the incident field, reflected field and transmitted field in this case are as follows,

$$\begin{cases} E_x^{\mathrm{i}} = E_0 \sin(k_0^a z - \omega_a t) - E_0 \sin(k_0^b z - \omega_b t) \\ H_y^{\mathrm{i}} = \sqrt{\varepsilon_0 / \mu_0}\, E_0 \sin(k_0^a z - \omega_a t) - \sqrt{\varepsilon_0 / \mu_0}\, E_0 \sin(k_0^b z - \omega_b t) \end{cases} \tag{21}$$

$$\begin{cases} E_x^{\mathrm{r}} = r_a E_0 \sin(k_0^a z - \omega_a t) - r_b E_0 \sin(k_0^b z - \omega_b t) \\ H_y^{\mathrm{r}} = -\sqrt{\varepsilon_0 / \mu_0}\, r_a E_0 \sin(k_0^a z - \omega_a t) + \sqrt{\varepsilon_0 / \mu_0}\, r_b E_0 \sin(k_0^b z - \omega_b t) \end{cases} \tag{22}$$

$$\begin{cases} E_x^{\mathrm{t}} = t_a E_0 \sin(k_a z - \omega_a t) - t_b E_0 \sin(k_b z - \omega_b t) \\ H_y^{\mathrm{t}} = \sqrt{\varepsilon_0 \varepsilon_{\mathrm{r}}^a / \mu_0 \mu_{\mathrm{r}}^a}\, t_a E_0 \sin(k_a z - \omega_a t) - \sqrt{\varepsilon_0 \varepsilon_{\mathrm{r}}^b / \mu_0 \mu_{\mathrm{r}}^b}\, t_b E_0 \sin(k_b z - \omega_b t) \end{cases} \tag{23}$$

Here, $r_\nu = \dfrac{1 - \sqrt{\varepsilon_{\mathrm{r}}^\nu / \mu_{\mathrm{r}}^\nu}}{1 + \sqrt{\varepsilon_{\mathrm{r}}^\nu / \mu_{\mathrm{r}}^\nu}}$ , $t_\nu = \dfrac{2}{1 + \sqrt{\varepsilon_{\mathrm{r}}^\nu / \mu_{\mathrm{r}}^\nu}}$ , and $k_\nu = -\sqrt{\varepsilon_{\mathrm{r}}^\nu \mu_{\mathrm{r}}^\nu}\, k_0^\nu$ are the reflection coefficients, transmission coefficients (for the electric field), and wave vector corresponding to $\omega_\nu$ with $\nu = a, b$ , respectively. $\varepsilon_{\mathrm{r}}^\nu$ and $\mu_{\mathrm{r}}^\nu$ are the relative permittivity and permeability corresponding to $\omega_\nu$ with $\nu = a, b$ , respectively. In the limit of $\omega_a \to \omega_b \to \omega$ , the reflection coefficient, transmission coefficient and wave vector corresponding to the central frequency $\omega$ will be denoted by $r_0$ , $t_0$ and $k$ , respectively.

Again, we focus on the momentum conversion process of this wave packet during a time interval [0, $\tau$] with $\tau = 2\pi / (\omega_a - \omega_b)$.

At $t$=0, the incident wave packet which locates at $z \in$ [-$c\tau$, 0] has the following momentum

$$P_{\text{vac}} = \int_{-c\tau}^{0} (g_{\text{em}})_{\text{in}} dz = \int_{-c\tau}^{0} \varepsilon_0 \mu_0 E_x^i H_y^i dz = \varepsilon_0 E_0^2 \tau \tag{24}$$

At $t$=$\tau$, part of this wave packet will penetrate into the metamaterial at the regime $z \in$ [0, $v_g\tau$], while another part of this wave packet will be reflected backward to the regime $z \in$ [-$c\tau$, 0]. It is noted that the group velocity $v_{\text{g}} \equiv \frac{\text{d}\omega}{\text{d}k} = \frac{\text{d}\omega}{\text{d}(-\omega\sqrt{\varepsilon_{\text{r}}\mu_{\text{r}}} / c)}$ is also positive in this case, since a normal dispersion, i.e., $\text{d}\varepsilon_{\text{r}} / \text{d}\omega > 0$, and $\text{d}\mu_{\text{r}} / \text{d}\omega > 0$ is assumed. The reflected wave packet which locates at $z \in$ [-$c\tau$, 0] has the following momentum

$$P_{\text{vac}2} = \int_{-c\tau}^{0} (g_{\text{em}})_{\text{reflection}} dz = \int_{-c\tau}^{0} \varepsilon_0 \mu_0 E_x^{\text{r}} H_y^{\text{r}} dz = -\varepsilon_0 E_0^2 r_0^2 \tau \tag{25}$$

The total momentum inside metamaterial consists of two parts, i.e., the electromagnetic part and the mechanical part. They are shown to have the following expressions (see Appendix)

$$P_{\text{em}} = \varepsilon_0 E_0^2 \tau \sqrt{\frac{\varepsilon_{\text{r}}}{\mu_{\text{r}}}} t_0^2 / \frac{\text{d}(-\sqrt{\varepsilon_{\text{r}}\mu_{\text{r}}}\omega)}{\text{d}\omega} \tag{26}$$

$$g_{\text{em}} = \frac{P_{\text{em}}}{v_{\text{g}}\tau} = \varepsilon_0 E_0^2 \sqrt{\frac{\varepsilon_{\text{r}}}{\mu_{\text{r}}}} t_0^2 / c \tag{27}$$

$$P_{\text{mech}} = \varepsilon_0 E_0^2 \tau t_0^2 \frac{1}{2} \left[ \frac{\varepsilon_{\text{r}}}{\mu_{\text{r}}} \frac{\text{d}}{\text{d}\omega} \left( \sqrt{\frac{\mu_{\text{r}}}{\varepsilon_{\text{r}}}} (\varepsilon_{\text{r}} - 1)\omega \right) + \frac{\text{d}}{\text{d}\omega} \left( \sqrt{\frac{\varepsilon_{\text{r}}}{\mu_{\text{r}}}} (\mu_{\text{r}} - 1)\omega \right) \right] / \frac{\text{d}(-\sqrt{\varepsilon_{\text{r}}\mu_{\text{r}}}\omega)}{\text{d}\omega} \tag{28}$$

$$g_{\text{mech}} = \frac{P_{\text{mech}}}{v_{\text{g}}\tau} = \varepsilon_0 E_0^2 t_0^2 \frac{1}{2} \left[ \frac{\varepsilon_{\text{r}}}{\mu_{\text{r}}} \frac{\text{d}}{\text{d}\omega} \left( \sqrt{\frac{\mu_{\text{r}}}{\varepsilon_{\text{r}}}} (\varepsilon_{\text{r}} - 1)\omega \right) + \frac{\text{d}}{\text{d}\omega} \left( \sqrt{\frac{\varepsilon_{\text{r}}}{\mu_{\text{r}}}} (\mu_{\text{r}} - 1)\omega \right) \right] / c \tag{29}$$

The above expressions (24)-(29) are the limiting results when $\omega_a \to \omega_b \to \omega$. Through some tedious derivation, we can still show that $P_{\text{vac}} = P_{\text{vac}2} + P_{\text{em}} + P_{\text{mech}}$ (see Appendix), which is the expression for momentum conservation in this situation (see Fig. 2). The mechanical momentum can be rewritten in a compact way,

$$P_{\text{mech}} = \varepsilon_0 E_0^2 \tau (1 + r_0^2 - t_0^2 \sqrt{\frac{\varepsilon_{\text{r}}}{\mu_{\text{r}}}} \frac{v_{\text{g}}}{c}) \tag{30}$$

$$g_{\text{mech}} = \varepsilon_0 E_0^2 (1 + r_0^2 - t_0^2 \sqrt{\frac{\varepsilon_{\text{r}}}{\mu_{\text{r}}}} \frac{v_{\text{g}}}{c}) / v_{\text{g}} \tag{31}$$

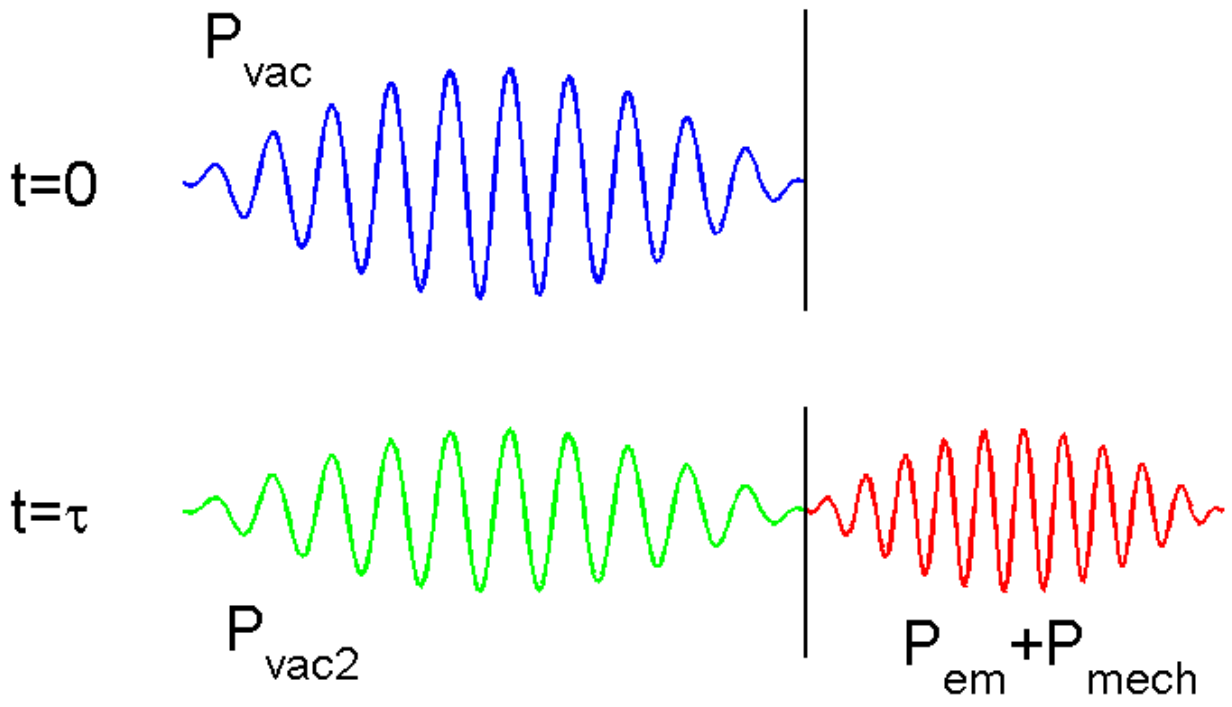

Figure 2. Illustration of momentum conservation at the interface between vacuum and an impedance-mismatched metamaterial. The incident wave packet (in blue) carries momentum $P_{\text{vac}}$, while the reflected wave (in green) carries momentum $P_{\text{vac2}}$, and the transmitted wave packet (in red) carries momentum $P_{\text{em}}+P_{\text{mech}}$.

It is shown that the mechanical momentum is in the same direction as energy flow, for

$$
\begin{aligned}
P_{\text{mech}} &= \varepsilon_0 E_0^2 \tau (1 + r_0^2 - t_0^2 \sqrt{\frac{\varepsilon_{\text{r}}}{\mu_{\text{r}}}} \frac{v_{\text{g}}}{c}) \\
&> \varepsilon_0 E_0^2 \tau (1 + r_0^2 - t_0^2 \sqrt{\frac{\varepsilon_{\text{r}}}{\mu_{\text{r}}}}) \\
&= \varepsilon_0 E_0^2 \tau (t_0^2 \frac{1+\frac{\varepsilon_{\text{r}}}{\mu_{\text{r}}}}{2} - t_0^2 \sqrt{\frac{\varepsilon_{\text{r}}}{\mu_{\text{r}}}}) = \frac{\varepsilon_0 E_0^2 \tau t_0^2}{2} (1 - \sqrt{\frac{\varepsilon_{\text{r}}}{\mu_{\text{r}}}})^2 \geq 0
\end{aligned} \tag{32}
$$

It is noted that the total momentum flow in metamaterial is,

$$
p_z^{\text{tot}} = (P_{\text{em}} + P_{\text{mech}}) / \tau = (1 + r_0^2) \varepsilon_0 E_0^2 \tag{33}
$$

which is again consistent with momentum flow density tensor

$$
T_{zz}^{\text{tot}} = T_{zz}^{\text{em}} = \frac{1}{2} [\varepsilon_0 (E_x^{\text{t}})^2 + \mu_0 (H_y^{\text{t}})^2] = \frac{1}{2} \varepsilon_0 E_0^2 t_0^2 \left( 1 + \frac{\varepsilon_{\text{r}}}{\mu_{\text{r}}} \right) \tag{34}
$$

by the aid of $(1 + r_0^2) / t_0^2 = (1 + \varepsilon_{\text{r}} / \mu_{\text{r}}) / 2$.

Based on the above calculation, we again conclude that inside a metamaterial, both the electromagnetic momentum and the mechanical momentum are in the same direction as the energy flow, instead of as the wave vector. Comparing with the impedance matching case, we also conclude that the ratio of the electromagnetic momentum density to the mechanical momentum density depends on the impedance and group velocity of electromagnetic waves inside the metamaterial.

## 4. ENERGY DENSITY, ENERGY OF WAVE PACKET AND ENERGY CONSERVATION

In the previous section, we have utilized the formalism of Yaghjian [14], where the field variables are $\vec{E}$ and $\vec{H}$, to show that the law of momentum conservation is preserved when the electromagnetic wave is launched into the metamaterial. We can derive the Poynting-type theorem using the formalism of Yaghjian [14], which is given by

$$\frac{\partial}{\partial t}\left(\frac{1}{2}\varepsilon_0\vec{E}^2+\frac{1}{2}\mu_0\vec{H}^2\right)+\nabla\cdot\left(\vec{E}\times\vec{H}\right)=-\left(\frac{\partial\vec{P}}{\partial t}\cdot\vec{E}+\mu_0\frac{\partial\vec{M}}{\partial t}\cdot\vec{H}\right) \tag{35}$$

This means that the power density that leads to the mechanical energy of the induced (polarized) electric and magnetic currents is $(\partial\vec{P}/\partial t)\cdot\vec{E}+(\mu_0\partial\vec{M}/\partial t)\cdot\vec{H}$.

As $\varepsilon_0\vec{E}^2/2=\mu_0\vec{H}^2/2$ for the electromagnetic wave in vacuum, the incident electromagnetic energy density in vacuum is $w_{\text{vac}}=\varepsilon_0\vec{E}^2$, which can be expressed explicitly by

$$\begin{aligned} w_{\text{vac}} &= \varepsilon_0\left[E_0\sin(k_0^a z-\omega_a t)-E_0\sin(k_0^b z-\omega_b t)\right]^2 \\ &= \varepsilon_0 E_0^2-\varepsilon_0 E_0^2\cos\left((k_0^a-k_0^b)z-(\omega_a-\omega_b)t\right) \\ &\quad -\varepsilon_0 E_0^2\left\{\frac{\cos 2(k_0^a z-\omega_a t)}{2}+\frac{\cos 2(k_0^b z-\omega_b t)}{2}-\cos\left((k_0^a+k_0^b)z-(\omega_a+\omega_b)t\right)\right\} \end{aligned} \tag{36}$$

The electromagnetic energy of a wave packet which locates at $z\in[-c\tau,0]$ is defined by

$$\text{E}_{\text{vac}}=\int_{-c\tau}^{0}w_{\text{vac}}(z,t)dz \tag{37}$$

It should be noted that the one-packet spatial integral of the highly-oscillating terms [where $\cos 2(k_0^\nu z-\omega_\nu t)$ or $\cos\left((k_0^a+k_0^b)z-(\omega_a+\omega_b)t\right)$ are involved] would vanish when we take the limit of $\omega_a\to\omega_b\to\omega$, and that the wave-packet spatial integral of the slowly-oscillating term $-\varepsilon_0 E_0^2\cos\left((k_0^a-k_0^b)z-(\omega_a-\omega_b)t\right)$ in $w_{\text{vac}}(z,t)$ is also zero, i.e.,

$$\begin{aligned} &-\varepsilon_0 E_0^2\int_{-c\tau}^{0}\cos\left((k_0^a-k_0^b)z-(\omega_a-\omega_b)t\right)dz \\ &= -\varepsilon_0 E_0^2\frac{1}{k_0^a-k_0^b}\left[\sin\left((k_0^a-k_0^b)(c\tau)+(\omega_a-\omega_b)t\right)-\sin\left((\omega_a-\omega_b)t\right)\right] \\ &= 0 \end{aligned} \tag{38}$$

Therefore, it follows that the incident electromagnetic energy of a wave packet is $\text{E}_{\text{vac}}=\varepsilon_0 E_0^2 c\tau$.

In the following part, we shall consider the problem of energy conservation during the electromagnetic interaction when the electromagnetic wave is launched normally into the metamaterial. We will address two cases, namely, impedance matching and mismatching cases, to show that the energy in these processes is also conserved.

**4.1 Energy conservation in the case of matched impedance**

When the impedance of the metamaterial is matched with that of vacuum, the explicit expression for electromagnetic energy $\text{E}_{\text{em}}=\int_0^{v_g\tau}w_{\text{em}}(z,t)dz$ in the metamaterial can be obtained in a similar manner as for obtaining electromagnetic energy $\mathrm{E}_{\mathrm{vac}}$ in vacuum. It can be readily verified that the wave-packet spatial integral of the highly-oscillating terms involved in the electromagnetic energy density $w_{\mathrm{em}}(z,t)$ in the metamaterial vanishes under the condition of $\omega_a \to \omega_b \to \omega$ . Besides, the wave-packet integral of the slowly-oscillating term $-\varepsilon_0 E_0^2 \cos\left((k_a-k_b)z-(\omega_a-\omega_b)t\right)$ in $w_{\mathrm{em}}(z,t)$ can be expressed by

$$
\begin{aligned}
& -\varepsilon_0 E_0^2 \int_0^{v_{\mathrm{g}}\tau} \cos\left((k_a-k_b)z-(\omega_a-\omega_b)t\right) dz \\
&= -\varepsilon_0 E_0^2 \frac{1}{k_a-k_b} \sin\left((k_a-k_b)z-(\omega_a-\omega_b)t\right)\Big|_0^{v_{\mathrm{g}}\tau} \\
&= -\varepsilon_0 E_0^2 \frac{1}{k_a-k_b} \left[\sin\left((k_a-k_b)v_{\mathrm{g}}\tau-(\omega_a-\omega_b)t\right)-\sin\left(-(\omega_a-\omega_b)t\right)\right] \\
&= 0
\end{aligned}
\tag{39}
$$

Therefore, the electromagnetic energy of a wave packet propagating in the metamaterial is $\mathrm{E}_{\mathrm{em}} = \varepsilon_0 E_0^2 v_{\mathrm{g}} \tau$ .

Let us now consider the interaction energy, which is the mechanical energy acquired by the induced electric and magnetic currents. The energy transferred by the electric interaction, of which the power density is $\left(\partial\vec{P}/\partial t\right)\cdot\vec{E}$ , in one wave packet is

$$
W_{\mathrm{e}} = \int_0^{\tau}\left(\int_0^{v_{\mathrm{g}}t} \frac{\partial\vec{P}}{\partial t}\cdot\vec{E}dz\right)dt \tag{40}
$$

where the induced electric current density is given by

$$
\frac{\partial P}{\partial t} = -\omega_a(\varepsilon_{\mathrm{r}}^a-1)\varepsilon_0 E_0\cos(k_a z-\omega_a t)+\omega_b(\varepsilon_{\mathrm{r}}^b-1)\varepsilon_0 E_0\cos(k_b z-\omega_b t) \tag{41}
$$

The explicit expression for the electric power density is

$$
\begin{aligned}
\frac{\partial P}{\partial t}\cdot E &= \left(-\omega_a(\varepsilon_{\mathrm{r}}^a-1)\varepsilon_0 E_0\cos(k_a z-\omega_a t)+\omega_b(\varepsilon_{\mathrm{r}}^b-1)\varepsilon_0 E_0\cos(k_b z-\omega_b t)\right) \\
&\quad\times\left(E_0\sin(k_a z-\omega_a t)-E_0\sin(k_b z-\omega_b t)\right) \\
&= \varepsilon_0 E_0^2\left[\frac{-\omega_a(\varepsilon_{\mathrm{r}}^a-1)\sin 2(k_a z-\omega_a t)-\omega_b(\varepsilon_{\mathrm{r}}^b-1)\sin 2(k_b z-\omega_b t)}{2}\right] \\
&\quad+\varepsilon_0 E_0^2\left[\omega_a(\varepsilon_{\mathrm{r}}^a-1)\frac{\sin\left((k_a+k_b)z-(\omega_a+\omega_b)t\right)-\sin\left((k_a-k_b)z-(\omega_a-\omega_b)t\right)}{2}\right] \\
&\quad+\varepsilon_0 E_0^2\left[\omega_b(\varepsilon_{\mathrm{r}}^b-1)\frac{\sin\left((k_b+k_a)z-(\omega_b+\omega_a)t\right)-\sin\left((k_b-k_a)z-(\omega_b-\omega_a)t\right)}{2}\right]
\end{aligned}
\tag{42}
$$

It should be noticed that the wave-packet integral of the highly-oscillating terms in $\left(\partial\vec{P}/\partial t\right)\cdot\vec{E}$ vanishes when the limit process of $\omega_a \to \omega_b \to \omega$ is taken. Thus, the only interesting term in $\left(\partial\vec{P}/\partial t\right)\cdot\vec{E}$ that deserves consideration is the slowly-oscillating term

$$A(z,t)=-\frac{\varepsilon_0E_0^2}{2}\left[\omega_a(\varepsilon_{\mathrm{r}}^a-1)-\omega_b(\varepsilon_{\mathrm{r}}^b-1)\right]\sin\left((k_a-k_b)z-(\omega_a-\omega_b)t\right) \tag{43}$$

Thus, the acquired mechanical energy of the induced electric current can be rewritten as

$$\begin{aligned}
W_{\mathrm{e}}&=\int_0^\tau\left(\int_0^{v_{\mathrm{g}}t}A(z,t)dz\right)dt\\
&=-\frac{\varepsilon_0E_0^2}{2}\left[\omega_a(\varepsilon_{\mathrm{r}}^a-1)-\omega_b(\varepsilon_{\mathrm{r}}^b-1)\right]\int_0^\tau\left[\int_0^{v_{\mathrm{g}}t}\sin\left((k_a-k_b)z-(\omega_a-\omega_b)t\right)dz\right]dt\\
&=-\frac{\varepsilon_0E_0^2}{2}\left[\omega_a(\varepsilon_{\mathrm{r}}^a-1)-\omega_b(\varepsilon_{\mathrm{r}}^b-1)\right]\frac{-1}{k_a-k_b}\int_0^\tau\left[\cos\left((k_a-k_b)z-(\omega_a-\omega_b)t\right)|_0^{v_{\mathrm{g}}t}\right]dt\\
&=\frac{\varepsilon_0E_0^2}{2}\left[\omega_a(\varepsilon_{\mathrm{r}}^a-1)-\omega_b(\varepsilon_{\mathrm{r}}^b-1)\right]\frac{1}{k_a-k_b}\left\{\frac{\sin\left[\left((k_a-k_b)v_g-(\omega_a-\omega_b)\right)\tau\right]}{(k_a-k_b)v_g-(\omega_a-\omega_b)}-\frac{\sin\left((\omega_a-\omega_b)\tau\right)}{\omega_a-\omega_b}\right\}
\end{aligned} \tag{44}$$

As the group velocity of the electromagnetic wave is $v_{\mathrm{g}}=\left(\omega_a-\omega_b\right)/\left(k_a-k_b\right)$, the $\omega_a\to\omega_b\to\omega$ limit of the term $\sin\left[\left((k_a-k_b)v_{\mathrm{g}}-(\omega_a-\omega_b)\right)\tau\right]\Big/\left[(k_a-k_b)v_{\mathrm{g}}-(\omega_a-\omega_b)\right]$ is $\tau$. Thus, the mechanical energy transferred by the electric interaction reads

$$W_{\mathrm{e}}=\frac{\varepsilon_0E_0^2}{2}\left[\frac{d(\omega\varepsilon_{\mathrm{r}})}{dk}-\frac{d\omega}{dk}\right]\tau=\frac{\varepsilon_0E_0^2}{2}\left(c-v_{\mathrm{g}}\right)\tau=\frac{\varepsilon_0E_0^2}{2}\left(1-\frac{v_{\mathrm{g}}}{c}\right)c\tau \tag{45}$$

when the $\omega_a\to\omega_b\to\omega$ limit is taken into account.

We now consider the mechanical energy $W_{\mathrm{m}}$ transferred by the magnetic interaction $\mu_0\left(\partial\vec{M}/\partial t\right)\cdot\vec{H}$. For the present case of impedance matching, we have the relation $\mu_0\left(\partial\vec{M}/\partial t\right)\cdot\vec{H}=\left(\partial\vec{P}/\partial t\right)\cdot\vec{E}$, and this implies that $W_{\mathrm{m}}=W_{\mathrm{e}}$. Hence, the total mechanical energy $W_{\mathrm{mech}}$ caused by the electromagnetic interaction $\left(\partial\vec{P}/\partial t\right)\cdot\vec{E}+\mu_0\left(\partial\vec{M}/\partial t\right)\cdot\vec{H}$ in one wave packet (propagating in the metamaterial) is given by

$$W_{\mathrm{mech}}=W_{\mathrm{e}}+W_{\mathrm{m}}=\varepsilon_0E_0^2\left(1-\frac{v_{\mathrm{g}}}{c}\right)c\tau \tag{46}$$

It can be found that there exists a relation of $\mathrm{E}_{\mathrm{vac}}=\mathrm{E}_{\mathrm{em}}+W_{\mathrm{mech}}$, i.e., the incident electromagnetic energy in vacuum equals the sum of the mechanical energy of the induced electromagnetic currents and the electromagnetic energy of the wave propagating in the metamaterial.

### 4.2 Energy conservation in the case of mismatched impedance

The incident electromagnetic energy of one wave packet in vacuum is $\mathrm{E}_{\mathrm{vac}} = \varepsilon_0 E_0^2 c\tau$. Obviously, the electromagnetic energy of one wave packet reflected back into vacuum is $\mathrm{E}_{\mathrm{vac2}} = \varepsilon_0 E_0^2 r_0^2 c\tau$ . We will calculate the transmitted electromagnetic energy density $\varepsilon_0 \vec{E}^2/2 + \mu_0 \vec{H}^2/2$ in the metamaterial. The explicit expression for the electric energy density is given by

$$
\begin{aligned}
\frac{1}{2}\varepsilon_0 E_t^2 &= \frac{1}{2}\varepsilon_0 \left[ t_a E_0 \sin(k_a z - \omega_a t) - t_b E_0 \sin(k_b z - \omega_b t) \right]^2 \\
&= \frac{1}{4}\varepsilon_0 E_0^2 \left( t_a^2 + t_b^2 \right) - \frac{1}{2}\varepsilon_0 E_0^2 t_a t_b \cos\left( (k_a - k_b) z - (\omega_a - \omega_b) t \right) \\
&\quad - \frac{1}{2}\varepsilon_0 E_0^2 \left\{ \frac{\cos 2(k_a z - \omega_a t)}{2} t_a^2 + \frac{\cos 2(k_b z - \omega_b t)}{2} t_b^2 - t_a t_b \cos\left( (k_a + k_b) z - (\omega_a + \omega_b) t \right) \right\}.
\end{aligned}
\tag{47}
$$

In the same fashion as the previous analysis, the spatial integrals of both the highly- and slowly-oscillating terms in one wave packet vanish. Thus, the electric energy, $\mathrm{E}_{\mathrm{e}} = \int_0^{v_{\mathrm{g}}\tau} \frac{1}{2}\varepsilon_0 E_t^2 dz$ , of the wave propagating inside the metamaterial is

$$
\mathrm{E}_{\mathrm{e}} = \frac{1}{4}\varepsilon_0 E_0^2 \left( t_a^2 + t_b^2 \right) v_{\mathrm{g}} \tau \tag{48}
$$

On the other hand, the explicit expression for the magnetic energy density is found to be

$$
\begin{aligned}
\frac{1}{2}\mu_0 H_t^2 &= \frac{1}{2}\mu_0 \frac{1}{\eta_0^2} \left[ \frac{1}{\eta_a} t_a E_0 \sin(k_a z - \omega_a t) - \frac{1}{\eta_b} t_b E_0 \sin(k_b z - \omega_b t) \right]^2 \\
&= \frac{1}{4}\varepsilon_0 E_0^2 \left( \frac{1}{\eta_a^2} t_a^2 + \frac{1}{\eta_b^2} t_b^2 \right) - \frac{1}{2}\varepsilon_0 E_0^2 \frac{t_a t_b}{\eta_a \eta_b} \cos\left( (k_a - k_b) z - (\omega_a - \omega_b) t \right) \\
&\quad - \frac{1}{2}\varepsilon_0 E_0^2 \left\{ \frac{\cos 2(k_a z - \omega_a t)}{2} \frac{1}{\eta_a^2} t_a^2 + \frac{\cos 2(k_b z - \omega_b t)}{2} \frac{1}{\eta_b^2} t_b^2 - \frac{t_a t_b}{\eta_a \eta_b} \cos\left( (k_a + k_b) z - (\omega_a + \omega_b) t \right) \right\}
\end{aligned}
\tag{49}
$$

where $\eta_0 \equiv \sqrt{\frac{\mu_0}{\varepsilon_0}}$ is the vacuum impedance and $\eta_\nu \equiv \sqrt{\frac{\mu_{\mathrm{r}}^\nu}{\varepsilon_{\mathrm{r}}^\nu}}$ is the relative impedance corresponding to $\omega_\nu$, with $\nu = a, b$. In the limit of $\omega_a \to \omega_b \to \omega$, the relative impedance corresponding to the central frequency $\omega$ will be denoted by $\eta_{\mathrm{r}}$ . As the spatial integrals of both the highly- and slowly-oscillating terms in one wave packet vanish, the magnetic energy of a wave packet in the metamaterial is

$$
\mathrm{E}_{\mathrm{m}} = \int_0^{v_{\mathrm{g}}\tau} \frac{1}{2}\mu_0 H_t^2 dz = \frac{1}{4}\varepsilon_0 E_0^2 \left( \frac{1}{\eta_a^2} t_a^2 + \frac{1}{\eta_b^2} t_b^2 \right) v_{\mathrm{g}} \tau \tag{50}
$$

Therefore, the total electromagnetic energy (for one wave packet) in the metamaterial is expressed by

$$
\mathrm{E}_{\mathrm{em}} = \mathrm{E}_{\mathrm{e}} + \mathrm{E}_{\mathrm{m}} = \frac{1}{4}\varepsilon_0 E_0^2 \left[ \left( 1 + \frac{1}{\eta_a^2} \right) t_a^2 + \left( 1 + \frac{1}{\eta_b^2} \right) t_b^2 \right] v_{\mathrm{g}} \tau \tag{51}
$$

which can be rewritten as $\mathrm{E}_{\mathrm{em}} = \frac{1}{2}\varepsilon_0 E_0^2\left(1+\frac{1}{\eta_{\mathrm{r}}^2}\right)t_0^2 v_{\mathrm{g}}\tau$ if we take the limit of $\omega_a \to \omega_b \to \omega$.

Now we shall consider the mechanical energy acquired by the induced electric current. The power density of the electric interaction is

$$\begin{aligned}\frac{\partial P}{\partial t}\cdot E &= \varepsilon_0 E_0^2\left[\frac{-\omega_a(\varepsilon_{\mathrm{r}}^a-1)t_a^2\sin 2(k_a z-\omega_a t)-\omega_b(\varepsilon_{\mathrm{r}}^b-1)t_b^2\sin 2(k_b z-\omega_b t)}{2}\right]\\ &+\varepsilon_0 E_0^2\left[\omega_a(\varepsilon_{\mathrm{r}}^a-1)t_a t_b\,\frac{\sin\left((k_a+k_b)z-(\omega_a+\omega_b)t\right)-\sin\left((k_a-k_b)z-(\omega_a-\omega_b)t\right)}{2}\right]\\ &+\varepsilon_0 E_0^2\left[\omega_b(\varepsilon_{\mathrm{r}}^b-1)t_a t_b\,\frac{\sin\left((k_b+k_a)z-(\omega_b+\omega_a)t\right)-\sin\left((k_b-k_a)z-(\omega_b-\omega_a)t\right)}{2}\right]\end{aligned} \tag{52}$$

Note that we only need to consider the slowly-oscillating term

$$B(z,t) = -\frac{\varepsilon_0 E_0^2}{2}t_a t_b\left[\omega_a(\varepsilon_{\mathrm{r}}^a-1)-\omega_b(\varepsilon_{\mathrm{r}}^b-1)\right]\sin\left((k_a-k_b)z-(\omega_a-\omega_b)t\right)$$

Thus, the transferred energy, $W_{\mathrm{e}} = \int_0^\tau\left(\int_0^{v_{\mathrm{g}}t}\frac{\partial\vec{P}}{\partial t}\cdot\vec{E}dz\right)dt$, caused by the electric interaction can be explicitly expressed by

$$\begin{aligned}W_{\mathrm{e}} &= \int_0^\tau\left(\int_0^{v_{\mathrm{g}}t}B(z,t)dz\right)dt = -\frac{\varepsilon_0 E_0^2}{2}t_a t_b\left[\omega_a(\varepsilon_{\mathrm{r}}^a-1)-\omega_b(\varepsilon_{\mathrm{r}}^b-1)\right]\int_0^\tau\left[\int_0^{v_{\mathrm{g}}t}\sin\left((k_a-k_b)z-(\omega_a-\omega_b)t\right)dz\right]dt\\ &= \frac{\varepsilon_0 E_0^2}{2}t_a t_b\left[\omega_a(\varepsilon_{\mathrm{r}}^a-1)-\omega_b(\varepsilon_{\mathrm{r}}^b-1)\right]\frac{1}{k_a-k_b}\int_0^\tau\left[\cos\left((k_a-k_b)v_{\mathrm{g}}t-(\omega_a-\omega_b)t\right)-\cos\left(-(\omega_a-\omega_b)t\right)\right]dt\\ &= \frac{\varepsilon_0 E_0^2}{2}t_a t_b\left[\omega_a(\varepsilon_{\mathrm{r}}^a-1)-\omega_b(\varepsilon_{\mathrm{r}}^b-1)\right]\frac{1}{k_a-k_b}\left\{\frac{\sin\left[\left((k_a-k_b)v_{\mathrm{g}}-(\omega_a-\omega_b)\right)\tau\right]}{(k_a-k_b)v_{\mathrm{g}}-(\omega_a-\omega_b)}-\frac{\sin\left((\omega_a-\omega_b)\tau\right)}{\omega_a-\omega_b}\right\}\\ &= \frac{\varepsilon_0 E_0^2}{2}t_a t_b\left[\omega_a(\varepsilon_{\mathrm{r}}^a-1)-\omega_b(\varepsilon_{\mathrm{r}}^b-1)\right]\frac{1}{k_a-k_b}\tau.\end{aligned}$$

The $\omega_a \to \omega_b \to \omega$ limit of the above result reads

$$W_{\mathrm{e}} = \frac{\varepsilon_0 E_0^2}{2}t_0^2\left[\frac{\mathrm{d}(\omega\varepsilon_{\mathrm{r}})}{\mathrm{d}k}-\frac{\mathrm{d}\omega}{\mathrm{d}k}\right]\tau \tag{53}$$

which can be rewritten as

$$W_{\mathrm{e}} = \frac{\varepsilon_0 E_0^2}{2}t_0^2\left[\frac{\mathrm{d}\left(-\omega\sqrt{\mu_{\mathrm{r}}\varepsilon_{\mathrm{r}}}\,\frac{1}{\eta_{\mathrm{r}}}\right)}{\mathrm{d}k}-\frac{\mathrm{d}\omega}{\mathrm{d}k}\right]\tau = \frac{\varepsilon_0 E_0^2}{2}t_0^2\left[\frac{c}{\eta_{\mathrm{r}}}+\omega\sqrt{\mu_{\mathrm{r}}\varepsilon_{\mathrm{r}}}\,\frac{1}{\eta_{\mathrm{r}}^2}\frac{\mathrm{d}\eta_{\mathrm{r}}}{\mathrm{d}k}-\frac{\mathrm{d}\omega}{\mathrm{d}k}\right]\tau \tag{54}$$

Here, relations $k = -\omega\sqrt{\mu_{\mathrm{r}}\varepsilon_{\mathrm{r}}}/c$ and $\mathrm{d}\left(-\omega\sqrt{\mu_{\mathrm{r}}\varepsilon_{\mathrm{r}}}\right)/\mathrm{d}k = c$ have been used.

The energy density of the magnetic interaction is given by

$$\mu_0 \frac{\partial M}{\partial t}\cdot H = \frac{1}{2}\varepsilon_0 E_0^2\left[-\frac{1}{\eta_a^2}\omega_a(\mu_{\rm r}^a-1)t_a^2\sin 2(k_a z-\omega_a t)-\frac{1}{\eta_b^2}\omega_b(\mu_{\rm r}^b-1)t_b^2\sin 2(k_b z-\omega_b t)\right]$$
$$+\varepsilon_0 E_0^2\left[\omega_a(\mu_{\rm r}^a-1)\frac{t_a t_b}{\eta_a\eta_b}\frac{\sin\left((k_a+k_b)z-(\omega_a+\omega_b)t\right)-\sin\left((k_a-k_b)z-(\omega_a-\omega_b)t\right)}{2}\right]$$
$$+\varepsilon_0 E_0^2\left[\omega_b(\mu_{\rm r}^b-1)\frac{t_a t_b}{\eta_a\eta_b}\frac{\sin\left((k_b+k_a)z-(\omega_b+\omega_a)t\right)-\sin\left((k_b-k_a)z-(\omega_b-\omega_a)t\right)}{2}\right]$$

Similarly, the wave-packet integrals of all the highly-oscillating terms vanish, and only the slowly-oscillating term

$$C(z,t) = -\frac{\varepsilon_0 E_0^2}{2}\frac{t_a t_b}{\eta_a\eta_b}\left[\omega_a(\mu_{\rm r}^a-1)-\omega_b(\mu_{\rm r}^b-1)\right]\sin\left((k_a-k_b)z-(\omega_a-\omega_b)t\right)$$

will be integrated for obtaining the magnetic interaction energy. The result is as follows

$$\begin{aligned}
W_{\rm m} &= \int_0^\tau\left(\int_0^{v_{\rm g}t} C(z,t)dz\right)dt \\
&= -\frac{\varepsilon_0 E_0^2}{2}\frac{t_a t_b}{\eta_a\eta_b}\left[\omega_a(\mu_{\rm r}^a-1)-\omega_b(\mu_{\rm r}^b-1)\right]\int_0^\tau\left[\int_0^{v_{\rm g}t}\sin\left((k_a-k_b)z-(\omega_a-\omega_b)t\right)dz\right]dt \\
&= \frac{\varepsilon_0 E_0^2}{2}\frac{t_a t_b}{\eta_a\eta_b}\left[\omega_a(\mu_{\rm r}^a-1)-\omega_b(\mu_{\rm r}^b-1)\right]\frac{1}{k_a-k_b}\left\{\frac{\sin\left[\left((k_a-k_b)v_{\rm g}-(\omega_a-\omega_b)\right)\tau\right]}{(k_a-k_b)v_{\rm g}-(\omega_a-\omega_b)}-\frac{\sin\left((\omega_a-\omega_b)\tau\right)}{\omega_a-\omega_b}\right\} \\
&= \frac{\varepsilon_0 E_0^2}{2}\frac{t_a t_b}{\eta_a\eta_b}\left[\omega_a(\mu_{\rm r}^a-1)-\omega_b(\mu_{\rm r}^b-1)\right]\frac{1}{k_a-k_b}\tau
\end{aligned} \tag{55}$$

The $\omega_a \to \omega_b \to \omega$ limit of the above formula gives $W_{\rm m} = \frac{\varepsilon_0 E_0^2}{2}\frac{t_0^2}{\eta_{\rm r}^2}\left[\frac{{\rm d}(\omega\mu_{\rm r})}{{\rm d}k}-\frac{{\rm d}\omega}{{\rm d}k}\right]\tau$. It can be rewritten as

$$W_{\rm m} = \frac{\varepsilon_0 E_0^2}{2}\frac{t_0^2}{\eta_{\rm r}^2}\left[\frac{{\rm d}\left(-\omega\sqrt{\mu_{\rm r}\varepsilon_{\rm r}}\eta_{\rm r}\right)}{{\rm d}k}-\frac{{\rm d}\omega}{{\rm d}k}\right]\tau = \frac{\varepsilon_0 E_0^2}{2}t_0^2\left[\frac{c}{\eta_{\rm r}}-\omega\sqrt{\mu_{\rm r}\varepsilon_{\rm r}}\frac{1}{\eta_{\rm r}^2}\frac{{\rm d}\eta_{\rm r}}{{\rm d}k}-\frac{1}{\eta_{\rm r}^2}\frac{{\rm d}\omega}{{\rm d}k}\right]\tau \tag{56}$$

Hence, the total mechanical energy transferred from the electromagnetic wave to the induced electromagnetic currents is

$$W_{\rm mech} = W_{\rm e} + W_{\rm m} = \varepsilon_0 E_0^2 t_0^2\left[\frac{1}{\eta_{\rm r}}-\frac{1}{2}\left(1+\frac{1}{\eta_{\rm r}^2}\right)\frac{v_{\rm g}}{c}\right]c\tau \tag{57}$$

Now we have obtained the incident electromagnetic energy $\mathrm{E}_{\rm vac} = \varepsilon_0 E_0^2 c\tau$ in vacuum, the reflected electromagnetic energy $\mathrm{E}_{\rm vac2} = \varepsilon_0 E_0^2 r_0^2 c\tau$, and the propagating electromagnetic energy $\mathrm{E}_{\rm em} = \frac{1}{2}\varepsilon_0 E_0^2\left(1+1/\eta_{\rm r}^2\right)t_0^2 v_{\rm g}\tau$ in the metamaterial. By aid of relation $r_0^2 + t_0^2/\eta_{\rm r} = 1$, one can find that the law of energy conservation, $\mathrm{E}_{\rm vac} = \mathrm{E}_{\rm vac2} + \mathrm{E}_{\rm em} + W_{\rm mech}$, is preserved in this process.

## 5. DISCUSSION AND CONCLUSION

The momentum of an electromagnetic wave in a medium has been studied for over a hundred years, and yet contradictive conclusions have been obtained by different approaches. Concerning the more complicated case of negative index metamaterial, which is not only polarized and magnetized, but also intrinsically dispersive, some attempts have been made to study the momentum of electromagnetic waves inside the metamaterial. In [20], the incompatibility between the photon momentum flow and the energy transfer for electromagnetic waves in a metamaterial has been pointed out. The inconsistency is supposed to stem from particle-wave duality [20]. Another formalism, based on classical electromagnetic theory, has been developed in [19]. The definitions of momentum density, momentum stress tensor and Lorentz force, are different from those in our present paper. Especially, the Lorentz force vanishes inside a lossless medium according to their formalism, and the force arising from the sudden change of electromagnetic stress tensor at the interface between vacuum and the metamaterial lacks a physical explanation [19].

In contrast, the present formalism provides a clear illustration of the electromagnetic momentum and the mechanical momentum, by defining the Lorentz force at the very beginning. Momentum conversion from the electromagnetic part into the mechanical part has been clearly demonstrated. The calculated results are consistent with the law of momentum conservation. Our additional calculation on energy flow shows that our formalism also obeys the law of energy conservation.

In summary, we have developed a theoretical approach for momentum in dispersive media with huge mass density. A clear identification of momentum density and momentum flow, for both the electromagnetic part and the mechanical part, has been obtained by implementing this formalism for a lossless dispersive metamaterial. Contrary to pervious studies on this subject [19, 20], our results show that the momentum density and momentum flow inside a metamaterial definitely have the same direction as the energy flow, rather than as the wave vector. We expect that our formalism may pave the way for easing the long-lasting dispute, or solving the incompatibility proposed recently in [20].

## APPENDIX

Here we show the calculation of mechanical momentum and the proof of momentum conservation for the case of mismatched impedance. The impedance matching case can be regarded as a special case of impedance mismatching.

**(1) Calculation of mechanical momentum for the case of mismatched impedance**

For the case of normal incidence, the generalized force is simplified to

$$\vec{f}_{\text{mech}} = \mu_0(\partial \vec{P}/\partial t)\times\vec{H} - \mu_0\varepsilon_0(\partial\vec{M}/\partial t)\times\vec{E} \tag{58}$$

Now we define $f_{\text{e}} = \mu_0(\partial \vec{P}/\partial t)\times\vec{H}$ , $f_{\text{h}} = -\mu_0\varepsilon_0(\partial\vec{M}/\partial t)\times\vec{E}$ , then the mechanical momentum acquired by the induced electric current is given by

$$P_{\mathrm{e}}=\int_0^{\tau}\int_0^{v_{\mathrm{g}}t} f_{\mathrm{e}}(z,t)dzdt$$

$$=\frac{\varepsilon_0 E_0^2}{c}\int_0^{\tau}\int_0^{v_{\mathrm{g}}t}\left\{\begin{array}{l}\dfrac{\partial}{\partial t}\left[(\varepsilon_{\mathrm{r}}^a-1)t_a\sin(k_a z-\omega_a t)-(\varepsilon_{\mathrm{r}}^b-1)t_b\sin(k_b z-\omega_b t)\right]\\ \times\left[t_a\sqrt{\dfrac{\varepsilon_{\mathrm{r}}^a}{\mu_{\mathrm{r}}^a}}\sin(k_a z-\omega_a t)-t_b\sqrt{\dfrac{\varepsilon_{\mathrm{r}}^b}{\mu_{\mathrm{r}}^b}}\sin(k_b z-\omega_b t)\right]\end{array}\right\}dzdt$$

$$=\frac{\varepsilon_0 E_0^2}{c}\int_0^{\tau}\int_0^{v_{\mathrm{g}}t}\left\{\begin{array}{l}\left[(-\omega_a)(\varepsilon_{\mathrm{r}}^a-1)t_a\cos(k_a z-\omega_a t)-(-\omega_b)(\varepsilon_{\mathrm{r}}^b-1)t_b\cos(k_b z-\omega_b t)\right]\\ \times\left[t_a\sqrt{\dfrac{\varepsilon_{\mathrm{r}}^a}{\mu_{\mathrm{r}}^a}}\sin(k_a z-\omega_a t)-t_b\sqrt{\dfrac{\varepsilon_{\mathrm{r}}^b}{\mu_{\mathrm{r}}^b}}\sin(k_b z-\omega_b t)\right]\end{array}\right\}dzdt$$

$$=\frac{\varepsilon_0 E_0^2}{c}\int_0^{\tau}\int_0^{v_{\mathrm{g}}t}\left\{\begin{array}{l}(-\omega_a)(\varepsilon_{\mathrm{r}}^a-1)t_a\cos(k_a z-\omega_a t)t_a\sqrt{\dfrac{\varepsilon_{\mathrm{r}}^a}{\mu_{\mathrm{r}}^a}}\sin(k_a z-\omega_a t)\\ +\omega_a(\varepsilon_{\mathrm{r}}^a-1)t_a\cos(k_a z-\omega_a t)t_b\sqrt{\dfrac{\varepsilon_{\mathrm{r}}^b}{\mu_{\mathrm{r}}^b}}\sin(k_b z-\omega_b t)\\ +\omega_b(\varepsilon_{\mathrm{r}}^b-1)t_b\cos(k_b z-\omega_b t)t_a\sqrt{\dfrac{\varepsilon_{\mathrm{r}}^a}{\mu_{\mathrm{r}}^a}}\sin(k_a z-\omega_a t)\\ -\omega_b(\varepsilon_{\mathrm{r}}^b-1)t_b\cos(k_b z-\omega_b t)t_b\sqrt{\dfrac{\varepsilon_{\mathrm{r}}^b}{\mu_{\mathrm{r}}^b}}\sin(k_b z-\omega_b t)\end{array}\right\}dzdt$$

$$=\frac{\varepsilon_0 E_0^2}{c}\int_0^{\tau}\int_0^{v_{\mathrm{g}}t}\left\{\begin{array}{l}(-\omega_a)(\varepsilon_{\mathrm{r}}^a-1)t_a^2\sqrt{\dfrac{\varepsilon_{\mathrm{r}}^a}{\mu_{\mathrm{r}}^a}}\dfrac{1}{2}\sin(2k_a z-2\omega_a t)\\ +\omega_a(\varepsilon_{\mathrm{r}}^a-1)t_a t_b\sqrt{\dfrac{\varepsilon_{\mathrm{r}}^b}{\mu_{\mathrm{r}}^b}}\dfrac{1}{2}\left\{\sin\left[(k_b-k_a)z-(\omega_b-\omega_a)t\right]-\sin\left[(k_b+k_a)z+(\omega_b+\omega_a)t\right]\right\}\\ +\omega_b(\varepsilon_{\mathrm{r}}^b-1)t_b t_a\sqrt{\dfrac{\varepsilon_{\mathrm{r}}^a}{\mu_{\mathrm{r}}^a}}\dfrac{1}{2}\left\{\sin\left[(k_a-k_b)z-(\omega_a-\omega_b)t\right]-\sin\left[(k_a+k_b)z+(\omega_a+\omega_b)t\right]\right\}\\ -\omega_b(\varepsilon_{\mathrm{r}}^b-1)t_b^2\sqrt{\dfrac{\varepsilon_{\mathrm{r}}^b}{\mu_{\mathrm{r}}^b}}\dfrac{1}{2}\sin(2k_b z-2\omega_b t)\end{array}\right\}dzdt$$

After integration over the spatial coordinate $z$, we obtain

$$P_{\mathrm{e}}=\frac{\varepsilon_0 E_0^2}{c}\int_0^{\tau}\left\{\begin{array}{l}(-\omega_a)(\varepsilon_{\mathrm{r}}^a-1)t_a^2\sqrt{\dfrac{\varepsilon_{\mathrm{r}}^a}{\mu_{\mathrm{r}}^a}}\dfrac{1}{4k_a}\left[\cos(2\omega_a t)-\cos(2k_a v_{\mathrm{g}}t-2\omega_a t)\right]\\ +\omega_a(\varepsilon_{\mathrm{r}}^a-1)t_a t_b\sqrt{\dfrac{\varepsilon_{\mathrm{r}}^b}{\mu_{\mathrm{r}}^b}}\dfrac{1}{2}\left\{\begin{array}{l}\dfrac{\cos\left[(\omega_a+\omega_b)t\right]-\cos\left[(k_a+k_b)v_{\mathrm{g}}t-(\omega_a+\omega_b)t\right]}{k_a+k_b}\\ +\dfrac{\cos\left[(\omega_b-\omega_a)t\right]-\cos\left[(k_b-k_a)v_{\mathrm{g}}t-(\omega_b-\omega_a)t\right]}{k_b-k_a}\end{array}\right\}\\ +\omega_b(\varepsilon_{\mathrm{r}}^b-1)t_b t_a\sqrt{\dfrac{\varepsilon_{\mathrm{r}}^a}{\mu_{\mathrm{r}}^a}}\dfrac{1}{2}\left\{\begin{array}{l}\dfrac{\cos\left[(\omega_a+\omega_b)t\right]-\cos\left[(k_a+k_b)v_{\mathrm{g}}t-(\omega_a+\omega_b)t\right]}{k_a+k_b}\\ +\dfrac{\cos\left[(\omega_a-\omega_b)t\right]-\cos\left[(k_a-k_b)v_{\mathrm{g}}t-(\omega_a-\omega_b)t\right]}{k_a-k_b}\end{array}\right\}\\ -\omega_b(\varepsilon_{\mathrm{r}}^b-1)t_b^2\sqrt{\dfrac{\varepsilon_{\mathrm{r}}^b}{\mu_{\mathrm{r}}^b}}\dfrac{1}{4k_b}\left[\cos(2\omega_b t)-\cos(2k_b v_{\mathrm{g}}t-2\omega_b t)\right]\end{array}\right\}dt$$

It should be noted that, in the limit of $\omega_a \to \omega_b \to \omega$, all the above terms will vanish after integrating over the time interval $[0, \tau]$, except the term $\cos[(k_b - k_a)v_g t - (\omega_b - \omega_a)t]$, whose value is exactly one, given that $v_{\mathrm{g}} = \dfrac{\omega_b - \omega_a}{k_b - k_a}$. Thus, the mechanical momentum caused by the electric current interaction is

$$\begin{aligned} P_{\mathrm{e}} &= \frac{\varepsilon_0 E_0^2}{c} \frac{\tau}{2(k_a - k_b)} \left[ \omega_a (\varepsilon_{\mathrm{r}}^a - 1) \sqrt{\frac{\varepsilon_{\mathrm{r}}^b}{\mu_{\mathrm{r}}^b}} - \omega_b (\varepsilon_{\mathrm{r}}^b - 1) \sqrt{\frac{\varepsilon_{\mathrm{r}}^a}{\mu_{\mathrm{r}}^a}} \right] t_a t_b \\ &= \frac{\varepsilon_0 E_0^2}{c} \frac{\tau}{2(k_a - k_b)} \left[ \omega_a (\varepsilon_{\mathrm{r}}^a - 1) \sqrt{\frac{\mu_{\mathrm{r}}^a}{\varepsilon_{\mathrm{r}}^a}} - \omega_b (\varepsilon_{\mathrm{r}}^b - 1) \sqrt{\frac{\mu_{\mathrm{r}}^b}{\varepsilon_{\mathrm{r}}^b}} \right] \sqrt{\frac{\varepsilon_{\mathrm{r}}^a \varepsilon_{\mathrm{r}}^b}{\mu_{\mathrm{r}}^a \mu_{\mathrm{r}}^b}} t_a t_b \end{aligned} \tag{59}$$

In the limit $\omega_a \to \omega_b \to \omega$, we get

$$P_{\mathrm{e}} = \frac{\varepsilon_0 E_0^2 \tau}{2} t_0^2 \frac{\varepsilon_{\mathrm{r}}}{\mu_{\mathrm{r}}} \frac{\dfrac{\mathrm{d}}{\mathrm{d}\omega}\left( \omega(\varepsilon_{\mathrm{r}} - 1)\sqrt{\dfrac{\mu_{\mathrm{r}}}{\varepsilon_{\mathrm{r}}}} \right)}{\dfrac{\mathrm{d}(-\omega\sqrt{\varepsilon_{\mathrm{r}}\mu_{\mathrm{r}}})}{\mathrm{d}\omega}} \tag{60}$$

In a similar way, we obtain

$$P_{\mathrm{h}} = \int_0^\tau \int_0^{v_{\mathrm{g}} t} f_{\mathrm{h}}(z,t) dz dt = \frac{\varepsilon_0 E_0^2 \tau}{2} t_0^2 \frac{\dfrac{\mathrm{d}}{\mathrm{d}\omega}\left( \sqrt{\dfrac{\varepsilon_{\mathrm{r}}}{\mu_{\mathrm{r}}}} (\mu_{\mathrm{r}} - 1)\omega \right)}{\dfrac{\mathrm{d}(-\sqrt{\varepsilon_{\mathrm{r}}\mu_{\mathrm{r}}}\omega)}{\mathrm{d}\omega}} \tag{61}$$

Finally, we obtain the momentum expression (28).

**(2) The proof of $P_{\mathrm{vac}} = P_{\mathrm{vac2}} + P_{\mathrm{em}} + P_{\mathrm{mech}}$ for the case of mismatched impedance**

In Appendix (1), the expression for the momentum inside the metamaterial, Eq. (28), is decomposed into two parts,

$$P_{\mathrm{mech}} = P_{\mathrm{e}} + P_{\mathrm{m}} \tag{62}$$

where $P_{\mathrm{e}} = \dfrac{\varepsilon_0 E_0^2}{2c} \dfrac{t_0^2}{\eta_{\mathrm{r}}^2} \dfrac{\mathrm{d}\left[\omega(\varepsilon_{\mathrm{r}} - 1)\eta_{\mathrm{r}}\right]}{\mathrm{d}k} \tau$, $P_{\mathrm{m}} = \dfrac{\varepsilon_0 E_0^2}{2c} t_0^2 \dfrac{\mathrm{d}\left[\omega(\mu_{\mathrm{r}} - 1)/\eta_{\mathrm{r}}\right]}{\mathrm{d}k} \tau$ with $\eta_{\mathrm{r}} = \sqrt{\dfrac{\mu_{\mathrm{r}}}{\varepsilon_{\mathrm{r}}}}$ and $k = -\dfrac{\omega}{c}\sqrt{\varepsilon_{\mathrm{r}}\mu_{\mathrm{r}}}$. And the two parts can be further rewritten as,

$$P_{\mathrm{e}}=\frac{\varepsilon_0 E_0^2}{2c}\frac{t_0^2}{\eta_{\mathrm{r}}^2}\left[\frac{\mathrm{d}\left(-\omega\sqrt{\mu_{\mathrm{r}}\varepsilon_{\mathrm{r}}}\right)}{\mathrm{d}k}-\frac{\mathrm{d}(\omega\eta_{\mathrm{r}})}{\mathrm{d}k}\right]\tau$$

$$=\frac{\varepsilon_0 E_0^2}{2c}\frac{t_0^2}{\eta_{\mathrm{r}}^2}\left[\frac{\mathrm{d}\left(-\omega\sqrt{\mu_{\mathrm{r}}\varepsilon_{\mathrm{r}}}\right)}{\mathrm{d}k}-\frac{\mathrm{d}\omega}{\mathrm{d}k}\eta_{\mathrm{r}}-\omega\frac{\mathrm{d}\eta_{\mathrm{r}}}{\mathrm{d}k}\right]\tau \tag{63}$$

$$=\frac{\varepsilon_0 E_0^2}{2c}\frac{t_0^2}{\eta_{\mathrm{r}}^2}\left[c-v_{\mathrm{g}}\eta_{\mathrm{r}}-\omega\frac{\mathrm{d}\eta_{\mathrm{r}}}{\mathrm{d}k}\right]\tau$$

$$P_{\mathrm{m}}=\frac{\varepsilon_0 E_0^2}{2c}t_0^2\left[\frac{\mathrm{d}\left(-\omega\sqrt{\mu_{\mathrm{r}}\varepsilon_{\mathrm{r}}}\right)}{\mathrm{d}k}-\frac{\mathrm{d}\left(\omega/\eta_{\mathrm{r}}\right)}{\mathrm{d}k}\right]\tau$$

$$=\frac{\varepsilon_0 E_0^2}{2c}t_0^2\left[\frac{\mathrm{d}\left(-\omega\sqrt{\mu_{\mathrm{r}}\varepsilon_{\mathrm{r}}}\right)}{\mathrm{d}k}-\frac{\mathrm{d}\omega}{\mathrm{d}k}\frac{1}{\eta_{\mathrm{r}}}+\frac{\omega}{\eta_{\mathrm{r}}^2}\frac{\mathrm{d}\eta_{\mathrm{r}}}{\mathrm{d}k}\right]\tau \tag{64}$$

$$=\frac{\varepsilon_0 E_0^2}{2c}t_0^2\left[c-\frac{v_{\mathrm{g}}}{\eta_{\mathrm{r}}}+\frac{\omega}{\eta_{\mathrm{r}}^2}\frac{\mathrm{d}\eta_{\mathrm{r}}}{\mathrm{d}k}\right]\tau$$

Consequently, we have the total mechanical momentum

$$P_{\mathrm{mech}}=\frac{\varepsilon_0 E_0^2}{2c}t_0^2\left[(1+\frac{1}{\eta_{\mathrm{r}}^2})c-\frac{2v_{\mathrm{g}}}{\eta_{\mathrm{r}}}\right]\tau \tag{65}$$

As we have obtained the electromagnetic momentum of one wave packet in vacuum and the metamaterial, i.e., Eqs. (24), (25) and (26), which are rewritten here as,

$$P_{\mathrm{vac}}=\varepsilon_0 E_0^2\tau \tag{66}$$

$$P_{\mathrm{vac2}}=-\varepsilon_0 E_0^2 r_0^2\tau \tag{67}$$

$$P_{\mathrm{em}}=\varepsilon_0 E_0^2\tau t_0^2\frac{v_{\mathrm{g}}}{\eta_{\mathrm{r}}c} \tag{68}$$

By the aid of $(1+r_0^2)/t_0^2=(1+1/\eta_{\mathrm{r}}^2)/2$, we can show that

$$P_{\mathrm{vac}}=P_{\mathrm{vac2}}+P_{\mathrm{em}}+P_{\mathrm{mech}} \tag{69}$$

This means that the momentum is conserved in this process.

## ACKNOWLEDGMENT

This work is supported by the National Natural Science Foundation of China (No. 60990320, and No. 60990322), and Natural Science Foundation of Zhejiang Province under Grant No.Y6100280. We are also grateful for valuable discussion with Dr. Yi Jin and Dr. Linfang Shen. One of the authors (JQ Shen) is also grateful to the support of the Fundamental Research Funds for the Central Universities of China.

**REFERENCES**

1. Jackson, J.D., *Classical electrodynamics*. 1999, New York: Wiley.
2. Minkowski. *Nachr. Ges. Wiss. Gottn Math.-Phys. KI.*, Vol. 53, No. 1908.
3. Abraham. *M. Rend. Circ. Matem. Palermo*, Vol. 28, No. 1, 1909.
4. Jones, R.V. and J.C.S. Richards. "The Pressure of Radiation in a Refracting Medium," *Proceedings of the Royal Society of London. Series A, Mathematical and Physical Sciences*, Vol. 221, No. 1147, 480-498, 1954.
5. Gordon, J.P. "Radiation Forces and Momenta in Dielectric Media," *Physical Review A*, Vol. 8, No. 1, 14-21, 1973.
6. Peierls, R. "The Momentum of Light in a Refracting Medium," *Proceedings of the Royal Society of London. Series A, Mathematical and Physical Sciences*, Vol. 347, No. 1651, 475-491, 1976.
7. Nelson, D.F. "Momentum, pseudomomentum, and wave momentum: Toward resolving the Minkowski-Abraham controversy," *Physical Review A*, Vol. 44, No. 6, 3985-3966, 1991.
8. Loudon, R., L. Allen, and D.F. Nelson. "Propagation of electromagnetic energy and momentum through an absorbing dielectric," *Physical Review E*, Vol. 55, No. 1, 1071-1085, 1997.
9. Mansuripur, M. "Radiation pressure and the linear momentum of the electromagnetic field," *Opt. Express*, Vol. 12, No. 22, 5375-5401, 2004.
10. Kemp, B., T. Grzegorczyk, and J. Kong. "Ab initio study of the radiation pressure on dielectric and magnetic media," *Opt. Express*, Vol. 13, No. 23, 9280-9291, 2005.
11. Mansuripur, M. "Radiation pressure and the linear momentum of light in dispersive dielectric media," *Opt. Express*, Vol. 13, No. 6, 2245-2250, 2005.
12. Scalora, M., et al. "Radiation pressure of light pulses and conservation of linear momentum in dispersive media," *Physical Review E*, Vol. 73, No. 5, 056604, 2006.
13. Mansuripur, M. "Radiation pressure and the linear momentum of the electromagnetic field in magnetic media," *Opt. Express*, Vol. 15, No. 21, 13502-13518, 2007.
14. Yaghjian, A.D. "Internal Energy, Q-Energy, Poynting's Theorem, and the Stress Dyadic in Dispersive Material," *Antennas and Propagation, IEEE Transactions on*, Vol. 55, No. 6, 1495-1505, 2007.
15. Pfeifer, R.N.C., et al. "Colloquium: Momentum of an electromagnetic wave in dielectric media," *Reviews of Modern Physics*, Vol. 79, No. 4, 1197-1216, 2007.
16. Veselago, V.G. "The electrodynamics of substances with simultaneously negative values of permittivity and permeability," *Sov. Phys. Usp.*, Vol. 10, 509-514, 1968.
17. Shelby, R.A., D.R. Smith, and S. Schultz. "Experimental Verification of a Negative Index of Refraction," *Science*, Vol. 292, No. 5514, 77-79, 2001.
18. Pendry, J.B. "Negative Refraction Makes a Perfect Lens," *Physical Review Letters*, Vol. 85, No. 18, 3966-3969, 2000.
19. Kemp, B.A., J.A. Kong, and T.M. Grzegorczyk. "Reversal of wave momentum in isotropic left-handed media," *Physical Review A*, Vol. 75, No. 5, 053810, 2007.
20. Veselago, V.G. "Energy, linear momentum, and mass transfer by an electromagnetic wave in a negative-refraction medium," *Physics-Uspekhi*, Vol. 52, No. 6, 649-654, 2009.